\documentclass[twocolumn]{revtex4-1}

\usepackage[english]{babel}
\usepackage{graphicx}
\usepackage{color}
\usepackage{amssymb,amsmath}
\usepackage{xspace}
\usepackage{upgreek}
\usepackage{ulem}
\usepackage{psfrag}
\usepackage{filemod}
\usepackage{siunitx}
\newcommand{\eg}{e.g.\@\xspace}
\newcommand{\ie}{i.e.\@\xspace}
\renewcommand{\bm}[1]{\boldsymbol{\mathbf{#1}}}
\newcommand{\ud}{\mathrm{d}}
\newcommand{\bra}{\left\langle}
\newcommand{\ket}{\right\rangle}
\newcommand{\re}{\operatorname{Re}}
\newcommand{\im}{\operatorname{Im}}
\newcommand{\eq}[1]{Eq.~\eqref{eq:#1}}

\newcommand{\Eq}[1]{Equation~\eqref{eq:#1}}

\newcommand{\fig}[1]{Fig.~\ref{fig:#1}}
\newcommand{\Fig}[1]{Figure~\ref{fig:#1}}
\newcommand{\neff}{n_{\text{eff}}^H}
\newcommand{\reff}{r_{\text{eff}}^H}

\begin{document}

\title{Enhanced Absorption of Waves in Stealth Hyperuniform Disordered Media}

\author{Florian Bigourdan}
\affiliation{Institut Langevin, ESPCI Paris, CNRS, PSL University, 1 rue Jussieu, 75005 Paris, France}
\author{Romain Pierrat}
\email[Corresponding author:]{romain.pierrat@espci.fr}
\affiliation{Institut Langevin, ESPCI Paris, CNRS, PSL University, 1 rue Jussieu, 75005 Paris, France}
\author{R\'emi Carminati}
\email[Corresponding author:]{remi.carminati@espci.fr}
\affiliation{Institut Langevin, ESPCI Paris, CNRS, PSL University, 1 rue Jussieu, 75005 Paris, France}

\begin{abstract}
   We study the propagation of waves in a set of absorbing subwavelength scatterers positioned on a stealth hyperuniform
   point pattern. We show that spatial correlations in the disorder substantially enhance absorption compared to a fully
   disordered structure with the same density of scatterers. The non-resonant nature of the mechanism provides broad
   angular and spectral robustness. These results demonstrate the possibility to design low-density materials with
   blackbody-like absorption.
\end{abstract}

\maketitle

\section{Introduction}

Enhancing the absorption of waves is of paramount importance in a number of applications, including solar
heating~\cite{SARBU-2016}, photovoltaics~\cite{POLMAN-2016}, molecular spectroscopy and sensing~\cite{OSAWA-2006},
acoustic insulation~\cite{SAPOVAL-1997,LEROY-2015}, seismology~\cite{BRULE-2014}, or ocean
waves~\cite{ARXIV-BENZAOUIA-2018}.  In optics, perfect absorption of coherent light has been demonstrated long-ago on
designed periodic structures supporting surface plasmons~\cite{HUTLEY-1976}. The conception of metallic nanostructures
to enhance absorption has become a strategy to increase the efficiency of photovoltaic conversion or photodetection of
visible or near infrared light, taking advantage of the broadband nature of plasmon resonances. A drawback is the strong
absorption in the metal itself, which reduces the absorption enhancement in the region of interest (\eg, the active
semiconductor). Optimized nanostructures also require costly nanofabrication techniques, and can be highly sensitive to
imperfections. The concept of coherent perfect absorption (CPA) has been generalized, showing that total absorption at a
given frequency can be reached in any absorbing material provided that the incident wavefront has been spatially shaped
to match an absorption eigenmode~\cite{CAO-2010-1,CAO-2011}. Coupled to wavefront shaping techniques, CPA permits total
absorption of light in complex disordered media~\cite{CHONG-2011,HSU-2015,LIEW-2016}. An important difference
between Refs.~\cite{CAO-2010-1,CAO-2011} and \cite{CHONG-2011,HSU-2015,LIEW-2016} is that the former are based on
resonant absorption by an eigenmode at certain frequency. However, the latter are based on non-resonant coupling of
light into an eigenchannel that can occur over a continuous frequency range and its enhancement effect can be
broadband.  Nevertheless, resonant absorption in designed nanostructures and CPA both require a coherent (shaped or
unshaped) incident wavefront. This makes the absorption process sensitive to changes in the incident wave (direction,
polarization and to a lesser extend spectrum), or to a reduction of its degree of coherence,
thus limiting the range of targeted applications.

Alternatively, absorption of natural light in disordered materials has been put forward recently, mostly in the context
of photovoltaics~\cite{SARGENT-2012,MUPPARAPU-2015,KOMAN-2016,LEE-2017}. It has been known for long that structural
correlations (\eg, correlations in the positions of scatterers dispersed in a homogeneous medium) have a strong impact
on the scattering
properties~\cite{MAURICE-1957,MARET-1990,CAO-2003-1,SCHEFFOLD-2004,SEGEV-2011,CONLEY-2014,LESEUR-2016-1,FROUFE-PEREZ-2017}.
The possibility to increase absorption using correlated disorder has been demonstrated in thin films patterned with
holes~\cite{VYNCK-2012}, and put forward recently in particular
cases~\cite{LESEUR-2016,WANG-2018,LIU-2018,FLORESCU-2018}. Nevertheless, a general strategy to optimize disordered
materials in terms of their ability to absorb light (and more generally waves) over a broad spectral and angular range
is still missing.

Previous works were often guided by the idea of increasing the optical path
length~\cite{MUPPARAPU-2015,KOMAN-2016}, or based on an optimization process without targeting a particular physical
mechanism~\cite{CHONG-2011,HSU-2015,LIEW-2016,LEE-2017}. In this manuscript, we propose an alternative
strategy. First, we demonstrate the existence of an upper bound for the absorbed power, and discuss the underlying
physical picture. This picture naturally dictates a methodology to enhance absorption in a disordered medium by
suppressing scattering using structural correlations, and maximizing the absorbance in the resulting effective
homogeneous medium. Second, based on numerical simulations, we show that hyperuniform materials, a specific class of
correlated disordered materials, permit to reach an absorption level close to the predicted upper bound, with
substantial spectral and angular robustness. The results demonstrate the substantial impact of structural correlations
on absorption, offering the possibility to design low-density disordered materials with blackbody-like absorption.

\section{Upper bound for wave absorption in disordered media}

Upper bounds for the absorbed power in photonic materials have been discussed in previous studies. One of them applies
to materials made of discrete scatterers with a fixed position in space, the degree of freedom being the
polarizabilities of the individual scatterers~\cite{HUGONIN-2015}. Since in the present work we focus on the degrees of
freedom offered by the spatial distribution of scatterers, this upper bound is not directly applicable. Another upper bound
was derived for a medium with given material properties (dielectric function) and a tunable geometry~\cite{MILLER-2016}.
Since our objective is to define strategies to optimize the absorption in a disordered medium described statistically,
we propose another approach to derive an upper bound that directly applies to the statistically averaged absorbed power.
As we will see, this upper bound derivation gives insight for the
definition of strategies to reach this bound [\ie to optimize the level of absorption by designing the statistical
properties (spatial correlations) of the disordered medium].

Let us consider a disordered medium embedded in a volume $V$ with external surface $S$, illuminated by a monochromatic
incident electric field with complex amplitude $\bm{E}_0$. From Poynting's theorem, and writing the total field
$\bm{E}=\bm{E}_0+\bm{E}_s$ with $\bm{E}_s$ the scattered field, energy consevation can be written in the form
$P_e=P_a+P_s$~\cite{JACKSON-1962}. In this expression, $P_e=\re\int_V \bm{j}\cdot\bm{E}^*_0\ud^3r/2$ is the extinction
power, $\bm{j}$ being the induced current density in volume~$V$, $P_a=\re\int_V\bm{j}\cdot\bm{E}^*\ud^3r/2$ is the
absorbed power, and $P_s=\re\oint_S (\bm{E}_s\times \bm{H}^*_s)\cdot \bm{n}_S \ud^2r/2$ is the scattered power,
$\bm{H}_s$ being the scattered magnetic field and $\bm{n}_S$ the outward normal on $S$.  Since we are interested in a
statisical description of the disordered medium, we introduce the ensemble average denoted by $\bra \cdots \ket$, and
write the fields and current density as a sum of an average value and a fluctuating part: $\bm{X} = \bra \bm{X} \ket +
\delta \bm{X}$ with $\bm{X} \in \left \{\bm{E},\bm{E}_s,\bm{H}_s,\bm{j} \right \}$ (note that $\bra \delta \bm{X} \ket =
0$, and that $\bm{E}_0 = \bra \bm{E}_0 \ket$ since $\bm{E}_0$ is deterministic). Introducing these expansions into the
expressions of the extinction, absorbed and scattered powers, they can be cast in the form $\bra P\ket=\overline{P}+
\widetilde{P}$, where $\overline{P}$ is the component involving average quantities $\bra \bm{X} \ket$ and
$\widetilde{P}$ the component involving fluctuations $\delta \bm{X}$ (see App.~\ref{app1}). Noting that
$\bra P_e \ket= \overline{P}_e$, energy conservation becomes on average
$\overline{P}_e=\overline{P}_a+\widetilde{P}_a+\overline{P}_s+\widetilde{P}_s$.  An important result of multiple
scattering theory states that the averaged field obeys a wave equation in an effective homogeneous
medium~\cite{SHENG-1995,MONTAMBAUX-2007}. Energy conservation for the averaged field can be written $ \overline{P}_e=
\overline{P}_a+ \overline{P}_s$  (see App.~\ref{app1}).  Combining the two preceding equations leads to $
\widetilde{P}_a+ \widetilde{P}_s=0$. As a result, the averaged absorbed power can be written
\begin{equation}\label{eq:thm1}
   \bra P_a\ket=\overline{P}_a - \widetilde{P}_s.
\end{equation}
Note that $\widetilde{P}_s \geq 0$ since it can be reduced to the integration of $|\delta \bm{E}_s|^2$ over a
closed-surface encompassing the medium in the far field. For the sake of simplicity, we now assume that the medium is a
slab of finite thickness and illuminated by a plane-wave $\bm{E}_0$ which will be the situation of interest in the
following. We consider the volume $V$ to be a portion of the slab with cross section $\Sigma$.
In that simple case, the power $\overline{P}_a$ absorbed in the effective homogeneous medium (as seen by the
averaged field) cannot exceed the incident power $P_0=\int_{\Sigma}\bm{\Pi}_0\cdot \bm{n}_\Sigma \ud^2r$ where
$\bm{\Pi_0}$ is the incident Poynting vector and $\bm{n}_\Sigma$ the inward normal on $\Sigma$.  We finally have
\begin{equation}\label{eq:thm2}
   \bra P_a\ket \leq P_0 - \widetilde{P}_s.
\end{equation}
The derivation of this upper bound for the averaged absorbed power in a disordered medium embedded in a slab geometry is
the first result in this article.

This upper bound suggests three strategies to increase the absorbed power: (1) Inhibit $\widetilde{P}_s$ while keeping
$\overline{P}_a$ approximately constant, (2) for a fixed $\widetilde{P}_s$, maximize $\overline{P}_a$ as close as
possible to $P_0$, or (3) inhibit $\widetilde{P}_s$ and maximize $\overline{P}_a$. Strictly speaking, finding the
statistial classes of disorder that maximize the absorbed power is challenging. Indeed, $\overline{P}_a$ and
$\widetilde{P}_s$ depend on transport parameters such as the effective refractive index and the scattering and
absorption mean free paths, that have a complex dependence on the structural correlations of
disorder~\cite{VAN_TIGGELEN-1994}. In the following we will constrain the problem by considering media made of absorbing
discrete scatterers dispersed in a transparent background, and characterized statistically by the spatial correlation in
the positions of the scatterers. Furthermore, we will consider the special class of stealth hyperuniform disorder, that
is known to produce $\widetilde{P}_s\ll P_0$ for large wavelengths even in dense materials~\cite{LESEUR-2016-1}. In
particular, a key question that we will address below concerns the validity of the preceding inequality in the
presence of strong absorption.

\section{Wave absorption in hyperuniform materials}

Hyperuniform point patterns are such that the variance of the number of points within a sphere of radius $R$
increases slower than the average number of points when $R$ tends to infinity~\cite{TORQUATO-2003}.
For a set of $N$ points with positions $\bm{r}_j$,
this property is equivalent to saying that the structure factor
\begin{equation}
   S(\bm{q})=\frac{1}{N}\left|\sum_{j=1}^N\exp[i\bm{q}\cdot\bm{r}_j]\right|^2
\end{equation}
vanishes when $|\bm{q}|\to 0$.  First introduced for their interest in close packing processes~\cite{TORQUATO-2003},
hyperuniform distributions of scatterers (hereafter denoted by hyperuniform materials) also produce materials of
interest in wave physics, in particular due to their ability to produce bandgaps even in absence of
periodicity~\cite{TORQUATO-2009,CHAIKIN-2013,MULLER-2013,TORQUATO-2013,TORQUATO-2013-2,AMOAH-2015,FROUFE-PEREZ-2016}.
Stealth hyperuniform materials is a specific class for which $S(\bm{q})$ strictly vanishes on a domain $\Omega$ of
typical size $K$ around $\bm{q}=0$~\cite{TORQUATO-2004,TORQUATO-2018}. The size $K$ actually controls the degree of
spatial correlations. For large $K$, the system is very constrained, generating short and long-range order. For $K \to
0$, structural correlations are relaxed and the system tends to a fully disordered material. The degree of order in the
pattern is usually measured by the ratio $\chi=[K/(2\pi)]^d/(2d\rho)$, with $\rho$ the density of points and $d$ the
dimension of space. $\chi$ has to be understood as the number of constrained degrees of freedom (DOF) normalized by the
total number of DOF.  $\chi=0$ corresponds to an uncorrelated pattern (fully disordered structure without any
constraints), while $\chi=1$ characterizes a perfect crystal with infinite range correlations~\cite{TORQUATO-2004}.

In the single-scattering regime, the scattered intensity is directly proportionnal to the structure factor $S(\bm{q})$,
with $\bm{q} = \bm{k}_s -  \bm{k}_i$, $\bm{k}_s$ and $\bm{k}_i$ being the scattered and incident wavevectors.
A stealth hyperuniform material does not scatter light for wavelengths satisfying  $\lambda>8\pi/K$, which
corresponds to scattering wavevectors $\bm{q}$ lying in the domain $\Omega$~\cite{LESEUR-2016-1}. In the
multiple-scattering regime, transparency also holds for large $\lambda$, provided that the
effective scattering mean-free path $\ell_s^H$ is larger than the system size $L$~\cite{LESEUR-2016-1} (we use the
superscript $H$ for quantities characterizing the hyperuniform material). In this regime, we have $\widetilde{P}_s^H\ll
P_0$, leading us to the idea that hyperuniform materials made with absorbing scatterers could be used to optimize the
absorbed power, along the line denoted by strategy~(1) above.

\begin{figure}[!htb]
   \centering
   \includegraphics[width=0.8\linewidth]{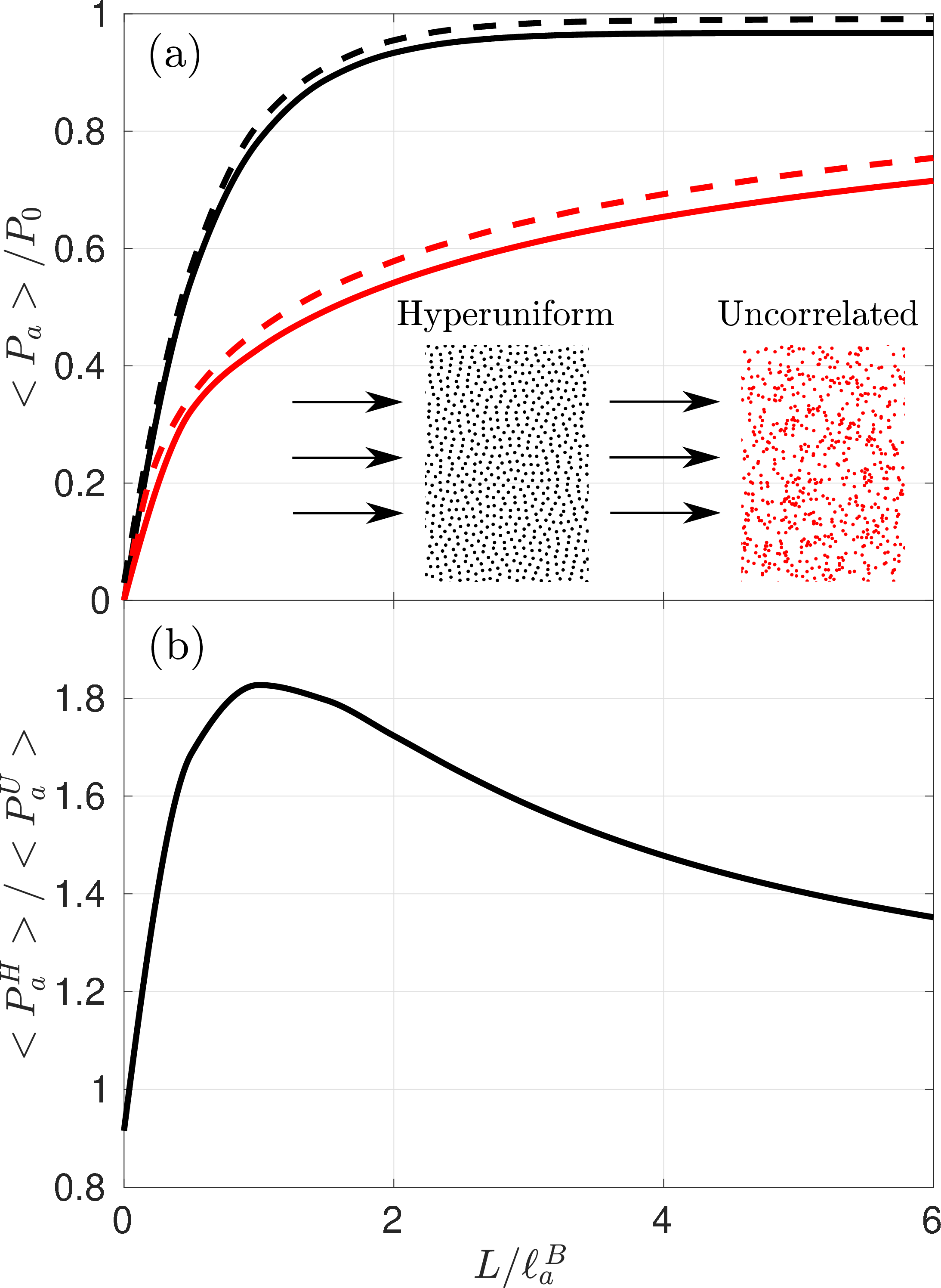}
   \caption{(a) Normalized absorbed power averaged over 60 configurations, versus the ISA absorption optical thickness
   $L/\ell_a^B$, for hyperuniform (black solid line) and uncorrelated (red solid line) materials. Parameters:
   $L/\ell_s^B=15$, $k_0\ell_s^B=13$, $N=30000$ scatterers, $\chi=0.44$. Black dotted line:  absorption by a homogeneous
   material with a complex refractive index equal to the effective index of the hyperuniform material (obtained by a
   fitting procedure). Red dotted line: absorption by an uncorrelated medium computed using a Monte Carlo
   simulation. (b) Ratio $\bra P_a^H \ket/\bra P_a^U \ket$ measuring the  absorption enhancement in the hyperuniform
   material.}
   \label{fig:absorbed_power}
\end{figure}

In order to support this idea, we present numerical simulations of wave scattering in two-dimensional model materials
made of subwavelength electric-dipole scatterers distributed on a stealth hyperuniform point pattern. To proceed, we
generate 60 configurations with thickness $L$ and transverse size $3L$, following the algorithm in
Ref.~\cite{TORQUATO-2004} (also described in Ref.~\cite{LESEUR-2016-1}). A value $\chi=0.44$ is chosen, permitting to
simulate large systems with a limited number of scatterers, although the behaviors discussed below are also observed for
smaller values of $\chi$. The value $\chi=0$ has also been chosen to make a comparison with a fully disordered
(uncorrelated) structure. In that case, the system is simply generated by picking-up random positions of the scatterers
uniformly in the volume $L\times 3L$. For simplicity we consider electromagnetic waves with an electric field
perpendicular to the plane containing the scatterers. Each scatterer is described by its electric polarizability
$\alpha(\omega)=-4\eta c^2/[Q(\omega^2-\omega_0^2)+i\omega^2]$.  Here $c$ is the speed of light in vacuum, $\omega_0$ is
the resonance frequency, $Q=\omega_0/\Gamma$ is the quality factor, $\Gamma=\Gamma_{\text{R}}+\Gamma_{\text{NR}}$ is the
linewidth, with a radiative and a non-radiative contribution (absorption), and $\eta=\Gamma_{\text{R}}/\Gamma$ is the
quantum efficiency (or albedo). From the polarizability, we can deduce the scattering cross-section
$\sigma_s=k_0^3|\alpha(\omega)|^2/4$, the extinction cross-section $\sigma_e=k_0\im[\alpha(\omega)]$ and the absorption
cross-section $\sigma_a = \sigma_e - \sigma_s$ of the scatterers, with $k_0=\omega/c=2\pi/\lambda$ the wavenumber in
vacuum. Given a number density $\rho$ of scatterers, we introduce as references the scattering and absorption mean-free
paths in the independent scattering approximation (ISA, also known as Boltzmann approximation), defined as
$\ell_s^B=(\rho\sigma_s)^{-1}$ and $\ell_a^B=(\rho\sigma_a)^{-1}$. The disordered medium is illuminated from the left by
a plane-wave at normal incidence (as shown schematically in the inset in Fig.~\ref{fig:absorbed_power}). Maxwell's
equations are solved using the coupled-dipoles method~\cite{LAX-1952}, and once the electric field illuminated each
scatterer is known, the total absorbed power $P_a^H$ is readily calculated (see App.~\ref{app2} for more details on the
numerical procedure). Repeating the calculation on the set of generated configurations, an ensemble average is performed
to obtain $\bra P_a^H \ket$. Running the same simulations on a set of configurations with uncorrelated scatterers, with
the same density $\rho$, leads to a calculation of the average power in a fully disordered material $\bra P_a^U\ket$ (we
use the superscript $U$ for quantities characterizing the uncorrelated material).

The dependence of $\bra P_a^H \ket$ and $\bra P_a^U \ket$ on the ISA absorption optical thickness $L/\ell_a^B$ is shown
in \fig{absorbed_power}\,(a). The parameter $L/\ell_a^B = \rho \sigma_a L$ is chosen here as a measure of the intrinsic
absorption level, independently of the spatial correlations in the medium. The power absorbed by the stealth hyperuniform
material is substantially larger than that absorbed in the fully disordered material with the same density. As shown in
\fig{absorbed_power}\,(b), we observe a maximum enhancement $\bra P_a^H \ket / \bra P_a^U \ket \simeq 1.8$ for a thickness
$L \simeq \ell_a^B$.

The propagation regimes in the hyperuniform and disordered structures can be analyzed, in order to get physical insight
into the mechanism responsible for the absorption enhancement. The effective scattering mean free path in the
non-absorbing hyperuniform structure is estimated to be $\ell_s^H\geq 750 \,\ell_s^B$, based on a computation of the
average field inside the medium (see App.~\ref{app3}). This confirms that hyperuniform correlations, with the set of
parameters chosen in \fig{absorbed_power}, lead to transparency by suppressing the scattered power~\cite{LESEUR-2016-1}.
Moreover, we have verified numerically that the fluctuations of the absorbed power also vanish (see App.~\ref{app4}), so
that error bars are not displayed in the figures for the sake of lisibility.  We can conclude that the hyperuniform
material in this regime behaves as an effective homogeneous and absorbing medium. Note that homogenization emerges here
as a consequence of spatial correlations, without any change in the density of scatterers. The descprition of the
effective medium has to go beyond standard approaches, such as the Maxwell-Garnett model.  The average absorbed power
can be calculated analytically considering a homogeneous slab, with a complex refractive index given by the effective
refractive index of the hyperuniform structure, also deduced from the computation of the average field inside the slab
(see App.~\ref{app5}).  The result of the analytical calculation is represented in  \fig{absorbed_power}\,(a), and
nicely fits the numerical simulation, confirming the picture of an homogenization process.  Regarding the uncorrelated
medium, scattering is not suppressed, but the absorption curve can be fairly reproduced using a Monte Carlo simulation
of intensity transport, describing the multiple scattering process in a medium with an effective refractive index
$n_{\text{eff}}^U=\sqrt{1+\rho\alpha(\omega)}$, and scattering and absorption mean-free paths $\ell_s^U=\ell_s^B$ and
$\ell_a^U=\ell_a^B$, as shown in \fig{absorbed_power}\,(a). We have also verified using the Monte Carlo simulation
that when $\ell_a^U\ll\ell_s^U$ the power absorbed by the uncorrelated structure approaches $P_0$. The reason is that
most of the light is absorbed before being scattered.

Interestingly, although spatial correlations in the stealth hyperuniform structures have a huge impact on the scattering
mean free path, they weakly affect the absorption mean free path. This result was already put forward in
Refs.~\cite{WANG-2018,LESEUR-2016} for hard spheres correlations, and is confirmed by numerical simulations for
hyperuniform disorder (see App.~\ref{app6}). Indeed, with the parameters used in \fig{absorbed_power}, we have
$\ell_a^H\simeq 0.7\ell_a^B$.  This calls for a simple random walk picture, that explains qualitatively the behavior of
$\bra P_a^H \ket$ and $\bra P_a^U \ket$. For the hyperuniform medium, we have $\ell_a^H\sim\ell_a^B<L\ll \ell_s^H$,
meaning that the photons travel along the distance $L$ without being scattered, with a high
probability to be absorbed before escaping. For the uncorrelated medium, $\ell_s^U\sim\ell_s^B<\ell_a^U\sim\ell_a^B<L$,
and a non-negligible fraction of photons are backscattered before being absorbed. These results are not specific to
the value of $\chi$ that has been chosen. Since the absorption mean-free path weakly depends on $\chi$, the results are valid
as soon as the transparent criterion is fulfilled~\cite{LESEUR-2016-1}.

\begin{figure}[!htb]
   \centering
   \includegraphics[width=0.80\linewidth]{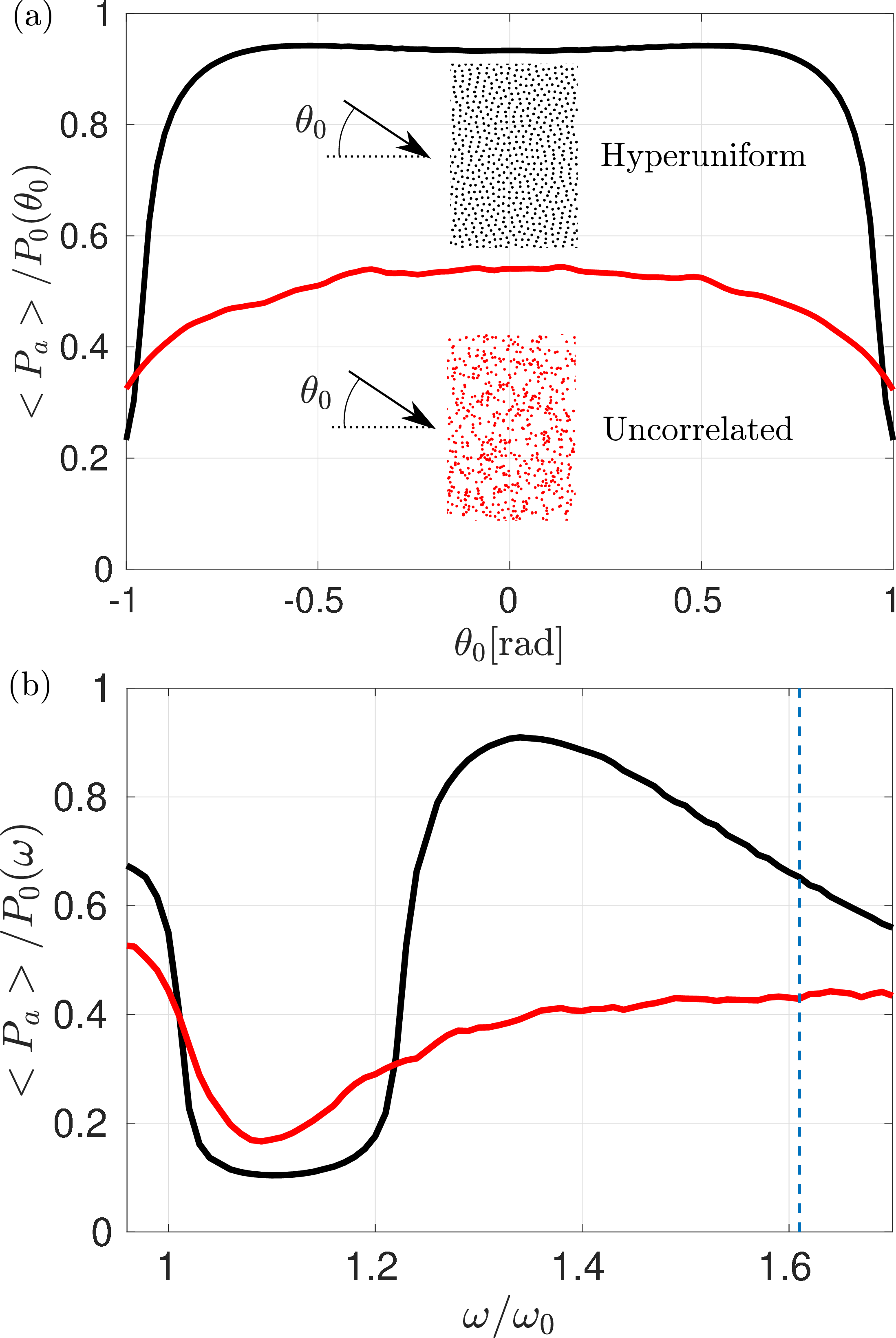}
   \caption{(a) Normalized average absorption versus the angle of incidence. Same parameters as in \fig{absorbed_power}
   with $L/\ell_a^B=2$.  (b) Normalized average absorption versus the illumination frequency at normal incidence.
   Same parameters as in \fig{absorbed_power} with $Q=3$ and $\eta=0.94$ which gives $L/\ell_s^B=15$,
   $L/\ell_a^B=1$ and $k_0\ell_s^B=13$ for $\omega/\omega_0=1.49$. The vertical dashed blue line indicates the cutoff
   frequency $\omega_c=Kc/4$ deliminating the transparency region $\lambda>8\pi/K$.}
   \label{fig:directional_spectral_robustness}
\end{figure}

From the validity of the intensity transport picture, we also expect the absorption enhancement to be robust against
changes in the direction of incidence and the illumination frequency. This is confirmed by the dependence of $\bra P_a
\ket$ on the incidence angle $\theta_0$ and the frequency $\omega$ displayed in \fig{directional_spectral_robustness}.
The dependence of $\bra P_a^H \ket$ on $\theta_0$ [\fig{directional_spectral_robustness}\,(a)] is weak (and weaker than
that observed for $\bra P_a^U \ket$), except at large angles where finite-size effects start to play an important role
(the transverse size of the medium is limited to $3L$ in the simulation). Using broadband scatterers (quality factor
$Q=3$), we obtain a large absorption enhancement on a broad frequency range
[\fig{directional_spectral_robustness}\,(b)].  The bandwidth with large absorption corresponds to the frequency range
for which the condition $\ell_a^H\sim\ell_a^B<L\ll \ell_s^H$ remains satisfied. It is interesting to note that the
disordered stealth hyperuniform material provides a maximum of absorption exceding that obtained with a periodic
crystal, together with a better angular robustness (numerical simulations for crystals are displayed in
App.~\ref{app2}).

Finally, it is interesting to use the analytical models to analyze the absorption enhancement at very large optical
thicknesses, that are not accessible to the full-wave numerical simulation. For the hyperuniform structure, we can
calculate the power absorbed in an effective homogeneous slab with thickness $L$. We assume a dilute medium in order to
neglect the index mismatch at the interfaces, and consider an absorption length $\ell_a^H\simeq\ell_a^B$ (see
App.~\ref{app6}). For the uncorrelated medium, we estimate the absorbed power using the diffusion
approximation valid for $L \gg \ell_s^B$ (see App.~\ref{app7}). Although not shown for brevity, the
enhancement curves have the same shape as that in \fig{absorbed_power}\,(b), with an enhancement reaching $\bra P_a^H \ket
/ \bra P_a^U \ket \simeq 7$ for $L=600\, \ell_s^B$ and $L\simeq \ell_a^B$.  We can even estimate the maximum enhancement
that can be expected using the analytical models (see App.~\ref{app8} for details on the calculation). We
obtain $(L/\ell_a^B)_{\text{optimum}}=1.256$ and
\begin{equation}\label{eq:max}
   \max\left[\frac{\bra P_a^H\ket}{\bra P_a^U\ket}\right]\simeq0.243\sqrt{\frac{L}{\ell_s^B}}.
\end{equation}

\section{Conclusion}

In summary, we have demonstrated theoretically the existence of an upper bound for the absorbed power in a disordered
medium, and analyzed the potential of hyperuniform materials for the design of strong absorbers of light. Hyperuniform
correlations can enhance the absorbed power very close to the upper bound limit, and the non-resonant nature of the
mechanism provides broad angular and spectral robustness. Robustness to positional variations of the scatterers is also
expected, since random displacements do not seem to destroy hyperuniformity~\cite{KIM-2018}.  Together with the
development of self-assembly processes able to produce hyperuniform architectures~\cite{RICOUVIER-2017,PIECHULLA-2018},
the results in this article could guide the design of disordered correlated materials with blackbody-like absorption.
The results should apply to all kinds of waves propagating in materials made of subwavelength absorbing scatterers in a
non-absorbing host medium.

\section*{Funding}

This research was supported by the ANR project LILAS under reference ANR-16-CE24-0001-01 and by LABEX WIFI (Laboratory
of Excellence ANR-10-LABX-24) within the French Program ``Investments for the Future'' under reference
ANR-10-IDEX-0001-02 PSL$^{\ast}$.

\appendix

\section{Elements of multiple scattering theory in disordered media}\label{app1}

Let us consider a random variable $\bm{X}$ for which a statistical average exists and is denoted by $\bra\bm{X}\ket$.
From that, we define the fluctuation $\delta\bm{X}$ for each statistical realization by the deviation to the average:
\begin{equation}
   \bm{X}=\bra\bm{X}\ket+\delta\bm{X}.
\end{equation}
In the case of disordered media, this expansion can be applied for example to the electric field:
\begin{equation}
   \bm{E}(\bm{r},\omega)=\bra \bm{E}(\bm{r},\omega)\ket+\delta \bm{E}(\bm{r},\omega)
\end{equation}
where the statistics is performed here over all possible spatial configurations of the disordered medium. From this
expansion, it is easy to compute the average intensity which reads
\begin{align}\nonumber
   \bra I(\bm{r},\omega)\ket = \bra\left|\bm{E}(\bm{r},\omega)\right|^2\ket
      & = \left|\bra \bm{E}(\bm{r},\omega)\ket\right|^2+\bra |\delta \bm{E}(\bm{r},\omega)|^2\ket
\\\label{eq:ballistic_diffuse_intensity}
      & = \overline{I}(\bm{r},\omega)+\widetilde{I}(\bm{r},\omega)
\end{align}
where we have used by definition $\bra \delta \bm{E}\ket=\bm{0}$. $\overline{I}$ and $\widetilde{I}$ are respectively
called the ballistic and the diffuse intensity. This expansion holds for any quadratic quantity such as the extinction,
absorbed and scattered powers used in the main text:
\begin{align}
   P_e(\omega) & = \frac{1}{2}\re\int_V \bm{j}\cdot\bm{E}^*_0\ud^3r,
\\
   P_a(\omega) & = \frac{1}{2}\re\int_V\bm{j}\cdot\bm{E}^*\ud^3r,
\\
   P_s(\omega) & = \frac{1}{2}\re\oint_S (\bm{E}_s\times \bm{H}^*_s)\cdot \bm{n}_S \ud^2r.
\end{align}
We thus have
\begin{align}
   \bra P_e(\omega)\ket & = \overline{P_e}(\omega)+\widetilde{P_e}(\omega) = \overline{P_e}(\omega),
\\
   \bra P_a(\omega)\ket & = \overline{P_a}(\omega)+\widetilde{P_a}(\omega),
\\
   \bra P_s(\omega)\ket & = \overline{P_s}(\omega)+\widetilde{P_s}(\omega)
\end{align}
where we have used $\bra \bm{E}_0\ket=\bm{E}_0$ and $\delta \bm{E}_0=\bm{0}$ since the incident field is deterministic.

\subsection{Ballistic beam}

From \eq{ballistic_diffuse_intensity}, we see that the ballistic intensity is directly given by the average field.
Thanks to the multiple scattering theory, we can show that this field propagates in a homogeneous medium with an
effective refractive index $n_{\text{eff}}(\omega)$~\cite{SHENG-1995,MONTAMBAUX-2007}. The imaginary part of this
refractive index gives the attenuation of the average field as it propagates in the medium. This attenuation is called
extinction and the typical distance over which the ballistic intensity is attenuated is the extinction mean-free path
$\ell_e$. We can also show that this attenuation is given by absorption (mean-free path $\ell_a$) and losses by scattering
(mean-free path $\ell_s$). We have the relation:
\begin{equation}
   \frac{1}{\ell_e}=\frac{1}{\ell_a}+\frac{1}{\ell_s}.
\end{equation}
The different powers relative to the effective medium read:
\begin{align}
   \overline{P_e}(\omega) & = \frac{1}{2}\re\int_V \bra\bm{j}\ket\cdot\bm{E}^*_0\ud^3r,
\\\label{eq:effective_medium_absorbed_power}
   \overline{P_a}(\omega) & = \frac{1}{2}\re\int_V\bra \bm{j}\ket\cdot\bra \bm{E}^*\ket\ud^3r,
\\
   \overline{P_s}(\omega) & = \frac{1}{2}\re\oint_S [\bra\bm{E}_s\ket\times \bra\bm{H}^*_s\ket]\cdot \bm{n}_S \ud^2r
\end{align}
where
\begin{multline}\label{eq:average_current}
   \bra \bm{j}(\bm{r},\omega)\ket=-i\omega\bra \bm{P}(\bm{r},\omega)\ket
\\
      =-i\omega\epsilon_0\left[n_{\text{eff}}(\omega)^2-1\right]\bra \bm{E}(\bm{r},\omega)\ket.
\end{multline}
We also have
\begin{equation}
   \overline{P_e}(\omega) = \overline{P_a}(\omega) + \overline{P_s}(\omega).
\end{equation}
\Eq{effective_medium_absorbed_power} together with \eq{average_current} are at the root of the computation of the
absorbed power in the stealth hyperuniform medium (see App.~\ref{app5}).

\subsection{Diffuse beam}

Regarding the computation of the diffuse intensity of \eq{ballistic_diffuse_intensity}, the problem is more complex. For
large optical thicknesses, we can show that it is governed by a diffusion equation which is at the root of the
computation of the absorbed power in the uncorrelated structure (see App.~\ref{app7}).

\section{Numerical computation of the average absorbed power}\label{app2}

\subsection{General method}

\begin{figure}[!htb]
   \centering
   \psfrag{a}[c]{(a)}
   \psfrag{b}[c]{(b)}
   \psfrag{TL}[c]{$3L$}
   \psfrag{L}[c]{$L$}
   \psfrag{t}[c]{$\theta_0$}
   \psfrag{V}[c]{$V$}
   \psfrag{S}[c]{$\Sigma$}
   \includegraphics[width=0.8\linewidth]{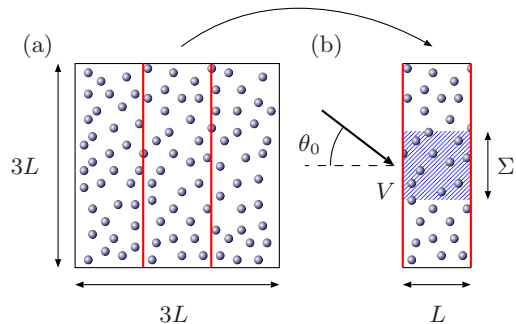}
   \caption{Illustration of the numerical computation process. (a) Configuration generated numerically on a square of
   size $3L$ containing $90000$ points. (b) One of the three bands used for the Maxwell simulation as a disordered
   pseudo-slab containing $30000$ scatterers. The absorbed power is computed on the blue hatched square $V$.
   $\theta_0$ is the angle of incidence of the plane-wave illumination.}
   \label{fig:numerical_process}
\end{figure}

The numerical computation is done through the following process schematized in \fig{numerical_process}. (1) First, we
generate $20$ disordered configurations containing $90000$ points in a square of size $3L$. For the hyperuniform point
patterns, we use the algorithm of Ref.~\cite{LESEUR-2016-1} and uncorrelated point patterns are simply generated
with a uniform random process. (2) To mimic a slab geometry and improve the statistical average, the generated
configurations are divided into three bands of thickness $L$ and length $3L$. After dressing all points with a
polarizability $\alpha(\omega)$, Maxwell's equations are solved in these configurations for a plane-wave illumination
from the left and using a coupled-dipoles method~\cite{LAX-1952}. In this formalism and for 2D scalar waves, the
exciting field on scatterer $i$ is given by
\begin{equation}\label{eq:coupled_dipoles}
   E_i(\omega)=E_0(\bm{r}_i,\omega)+k_0^2\alpha(\omega)\sum_{j=1,j\ne i}^NG_0(\bm{r}_i-\bm{r}_j)E_j(\omega)
\end{equation}
where $E_0$ is the incident field and $G_0$ is the Green function in vacuum. It connects the field to
its point dipole source and is given by
\begin{equation}
   G_0(\bm{r}-\bm{r}')=\frac{i}{4}\operatorname{H}_0^{(1)}(k_0|\bm{r}-\bm{r}'|)
\end{equation}
where $\operatorname{H}_0^{(1)}$ is the Hankel function of zero order and first kind. The computation of the exciting
fields consists in solving the linear system given by \eq{coupled_dipoles}. (3) Finally, to avoid finite-size
effects again, the absorbed power is computed only for scatterers belonging to the central square of volume $V$ using the
relation
\begin{equation}\label{eq:absorbed_power}
   \frac{P_a}{P_0}=\frac{\sigma_a}{L|E_0|^2}\sum_{i\in V}|E_i|^2
\end{equation}
and a statistical average is performed using the $60$ different realizations of the disorder to get $\bra P_a\ket/P_0$.

\subsection{Comparison with crystalline structures}

\begin{figure}[!htb]
   \centering
   \includegraphics[width=0.8\linewidth]{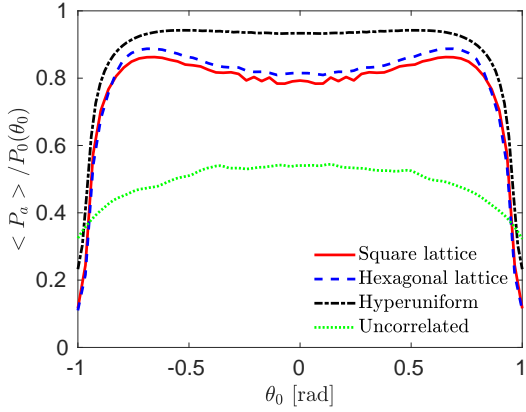}
   \caption{Comparison of the power absorbed by different types of structures as a function of the incident angle
   $\theta_0$. For all simulations, the geometry (slab), the number of scatterers and the optical properties of each
   scatterer are preserved. Parameters: $L/\ell_s^B=15$, $L/\ell_a^B=1$, $k_0\ell_s^B=13$ and $N=30000$ scatterers.}
   \label{fig:crystals}
\end{figure}

We have also performed numerical simulations in the cases of fully ordered structures to compared with the hyperuniform
and uncorrelated ones. The results are reported in \fig{crystals}. For the periodic lattices, the unit cell
parameter $a$ is such that $k_0a=1.95$ for the square lattice and $k_0a=2.10$ for the hexagonal lattice. This means that
in both cases $a<\lambda$ and there is only one propagating Bloch mode (no Bragg peaks are visible). However,
$\rho\lambda^2$ is not large enougn for the crystalline structures to be homogenized (in the usual sense) and described
by an effective refractive index. This makes the comparison to the hyperuniform structures difficult to address using
simple arguments.

\section{Average field evolution and effective refractive index fit for the stealth hyperuniform structure}\label{app3}

\begin{figure}[!htb]
   \centering
   \psfrag{E}[c]{$E_0$}
   \psfrag{n}[c]{$\neff$}
   \psfrag{O}[c]{$O$}
   \psfrag{L}[c]{$L$}
   \psfrag{z}[c]{$z$}
   \includegraphics[width=0.5\linewidth]{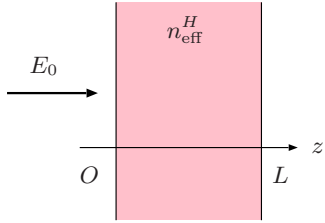}
   \caption{System of interest to compute the average field (homogeneous slab).}
   \label{fig:slab}
\end{figure}

To compute analytically the power absorbed by the stealth hyperuniform structures, we have to compute
the evolution of the average field which allows us also to estimate the effective refractive index $\neff$. The system of
interest is depicted in \fig{slab}. It is illuminated by a plane-wave at normal incidence given by
\begin{equation}
   E_0(z)=E_0\exp(ik_0z).
\end{equation}

First, we compute numerically the average field for several depths inside the slab. For that purpose, we compute for each
configuration the field at any position using the expression
\begin{equation}\label{eq:field}
   E(\bm{r},\omega)=E_0(\bm{r},\omega)+k_0^2\alpha(\omega)\sum_{j=1}^NG_0(\bm{r}-\bm{r}_j)E_j(\omega)
\end{equation}
and the expressions of the exciting fields given by \eq{coupled_dipoles}. Then
we perform the statistical average. The numerical result is finally fitted with a standard Fabry-Perot approach
which leads to the theoretical expression of the average field given by
\begin{multline}\label{eq:average_field}
   \bra E(z)\ket = \frac{t}{1-{\reff}^2}\left\{\exp[ik_0(\neff-1)z]
\right.\\\left.
      +\gamma\exp[-ik_0(\neff+1)z]\right\}E_0(z)
\end{multline}
where $\reff=r\exp(i\neff k_0L)$, $\gamma=r\exp(2i\neff k_0L)$,
$r=(\neff-1)/(\neff+1)$ is the Fresnel amplitude coefficient in reflection and
$t=2/(\neff+1)$ is its counterpart in transmission. The fit is done using an optimization algorithm,
based on a (global) genetic algorithm combined with a (local) Newton algorithm~\cite{SPALL-2003}. Note that in the most
general case, the effective refractive index is non-local but it can be very well approximated by a simple complex
constant for statistically homogeneous dilute systems.

\begin{figure*}[!htb]
   \centering
   \psfrag{a}[c]{(a)}
   \psfrag{b}[c]{(b)}
   \psfrag{E}[c][][1][90]{$\left|\bra E\ket\right|^2/I_0$}
   \psfrag{z}[c]{$k_0z$}
   \includegraphics[width=1.0\linewidth]{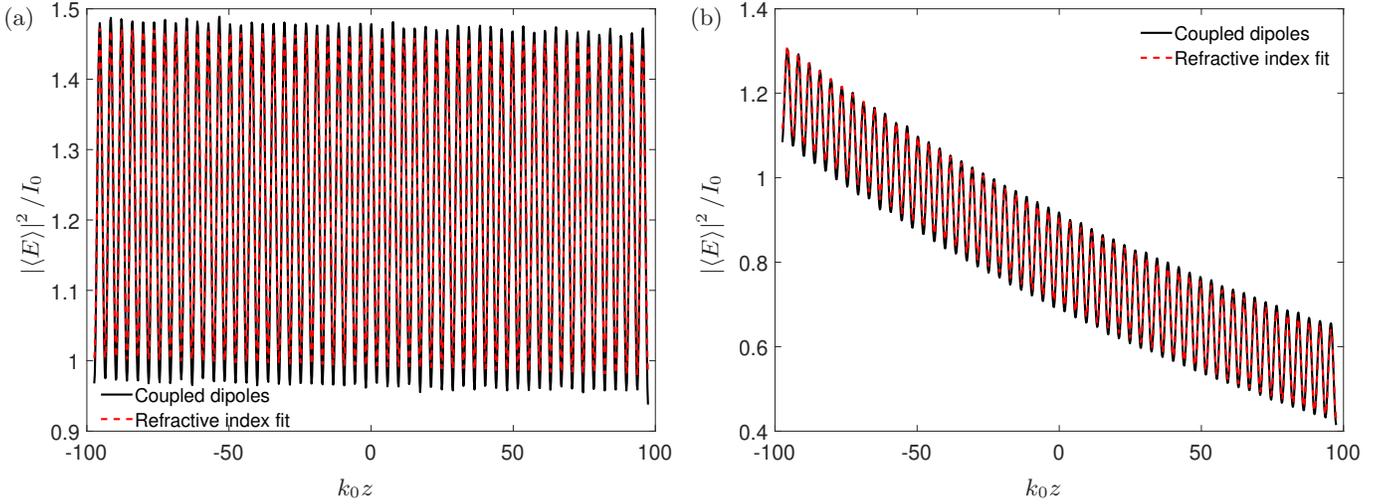}
   \caption{Example of fitting of the effective refractive index of the average field inside the cloud. The average
   field is compared with the analytical solution of the field propagating inside an (transversally) infinite
   homogeneous slab of same refractive index. Calculations were done on $60$ stealth hyperuniform configurations made of
   $N=30000$ dipoles with $\chi=0.444$, $L/\ell_s^B=15$ and (a) $L/\ell_a^B=0$ (no absorption) or (b) $L/\ell_a^B=0.5$.}
   \label{fig:average_field}
\end{figure*}

\Fig{average_field} gives two examples for a stealth hyperuniform structure. From the effective refractive index, we get
the extinction mean-free path by
\begin{equation}
   \ell_e^H=\frac{1}{2k_0{\neff}''}
\end{equation}
where ${\neff}''$ is the imaginary part of the effective refractive index.

In \fig{average_field}\,(a), there is no absorption and the extinction mean-free path can be assimilated to the
scattering mean-free path. We clearly see the very slow decay of the intensity of the average field. which means that
$\ell_s^H\gg L$ and the structure is transparent. The fit gives for \fig{average_field}\,(a)
$\ell_s^H\in[750,1200]\ell_s^B$. This large window results from large incertainties because of a small decay. On the
contrary, absorption is present in \fig{average_field}\,(b). In that case, we cannot discriminate between the scattering
and the absorption processes through the evolution of the average field. However, as pointed out in
App.~\ref{app4}, the fluctuations of the absorbed power vanish, which probably proves that $\ell_s^H$ remains
large compared to the size of the system $L$. Thus the imaginary part of the refractive index gives here a direct access
to the absorption length.  The fit gives for \fig{average_field}\,(b) $\ell_a^H\in[0.71,0.83]\ell_a^B$. The
proximity of $\ell_a^H$ with $\ell_a^B$ is surprising. It has already been put forward recently for a different
correlation type (hard-sphere potential)~\cite{LESEUR-2016,WANG-2018} and needs a more refined analysis that is left for
future work. In particular, \eq{absorbed_power} could be used as a starting point for the derivation of a theoretical
result regarding the expression of $\ell_a^H$.

\section{Standard deviation of the absorbed power distribution}\label{app4}

\begin{figure}[!htb]
   \centering
   \includegraphics[width=0.8\linewidth]{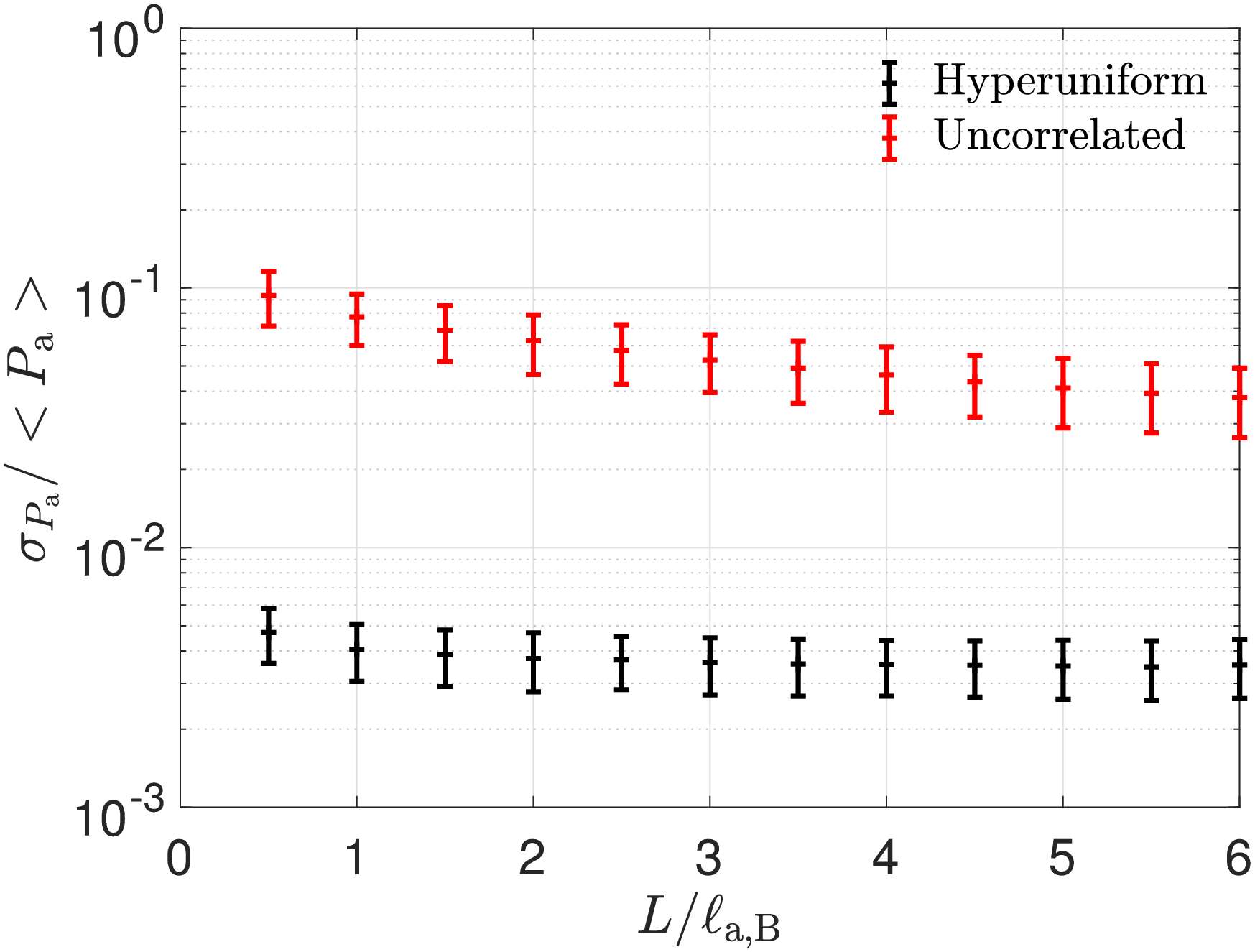}
   \caption{Relative standard deviation of the absorption  in a correlated (black, down) and uncorrelated (red, up)
   cloud. Same parameters as in Fig.~1 of the main text. Error bars were estimated by boostrapping on $10000$ resampling and
   taking $\pm 3$ times the standard deviation from the displayed value.}
   \label{fig:standard_deviation}
\end{figure}

To check the transparent character of the stealth hyperuniform structure for the set of parameters used in
this work, we can compare the standard deviation of the absorbed power distribution for both types of correlations
(hyperuniform and fully disordered system). \Fig{standard_deviation} shows the computed ensemble standard deviation of
the power absorbed $\sigma_{P_a}$ relative to the ensemble mean power absorbed $\bra P_a\ket$.  It indicates clearly the
strong reduction of $\sigma_{P_a}$ for a stealth hyperuniform medium in the transparency regime, compared to an
uncorrelated medium with the same intrinsic parameters, typically by one to two orders of magnitude. Furthermore, its
amplitude in the hyperuniform case is more than two orders of magnitude smaller than  $\bra P_a\ket$, which indicates
that $\bra P_a^H\ket \simeq \overline{P}_a^H$. The power associated to absorption by the average field suffices to
describe the absorption of a hyperuniform medium in the transparency regime for most applications. This relatively low
standard deviation of the power absorbed is interpreted by the fact that a stealth hyperuniform medium in the
transparency regime suppresses most of the ensemble fluctuations of the fields by definition of this regime. One may
then expect that all the observables are described by the average fields. Note also that in the case of uncorrelated
media, the larger the optical thickness $L$ compared to $\ell_s^U$, the closer $\widetilde{P}_s\rightarrow P_0$ and
typically the larger $\sigma_{P_a^U}$ relatively to $\bra P_a^U\ket$.

\section{Average absorbed power for the stealth hyperuniform structure}\label{app5}

Even in the presence of absorption, $\ell_s^H\gg L$ and the stealth hyperuniform structure remains transparent. This
means that the fluctuations of the field vanish and $\overline{P}_s^H\sim 0$ leading to $\bra
P_a^H\ket=\overline{P}_a^H$. Thus the power absorbed by the system is given by
\begin{equation}\label{eq:absorbed_power_hyper}
   \frac{\bra P_a^H\ket}{P_0}
      =\frac{\omega}{2}\int\im[\bra P\ket\bra E^*\ket]\ud z\left[\frac{\epsilon_0 c I_0}{2}\right]^{-1}
\end{equation}
where the average polarization field is given by $\bra P\ket=\epsilon_0[{\neff}^2-1]\bra E\ket$. Using \eq{average_field},
we easily obtain an analytical and exact expression of the power absorbed by the slab as a function of its size $k_0L$
and effective refractive index $\neff$.  In order to get more physical insights into the absorption in the slab,
approximations are possible. Indeed, if the slab is large enough, \ie ${\neff}''k_0L\gg 1$, round trips are not allowed
and a very good approximation of \eq{average_field} is given by
\begin{equation}
   \bra E(z)\ket = t\exp(ik_0\neff z)E_0.
\end{equation}
Using this expression in \eq{absorbed_power_hyper} leads to
\begin{multline}
   \frac{\bra P_a^H\ket}{P_0}=\frac{|t|^2}{2{\neff}''}\im\left[{\neff}^2\right]\left[1-\exp(-2k_0{\neff}'' L)\right]
\\
      =\left[1-R\right]\left[1-\exp(-2k_0{\neff}'' L)\right]
\end{multline}
where $R=|r|^2$ is the factor of reflectivity in intensity of a single interface. We finally have
\begin{equation}\label{eq:P_a_hyp}
   \frac{\bra P_a^H\ket}{P_0}=\left[1-R\right]\left[1-\exp\left(-\frac{L}{\ell_a^H}\right)\right]
\end{equation}
which is the expression used in Fig.~1\,(a) of the main text (with a fitted refractive index).

\section{Absorption mean-free path and hyperuniformity}\label{app6}

\begin{figure}[!htb]
   \centering
   \psfrag{L}[c]{$L/\ell_a^B$}
   \psfrag{P}[c]{$\bra P_a^H\ket/P_0$}
   \psfrag{p}[l]{$\bra P_a^H\ket/P_0=1-\exp[-1.4L/\ell_a^B]$}
   \psfrag{a}[l]{$k_0\ell_s^B=13,L=11\ell_s^B,\chi=0.22$}
   \psfrag{b}[l]{$k_0\ell_s^B=13,L=15\ell_s^B,\chi=0.44$}
   \psfrag{c}[l]{$k_0\ell_s^B=18,L=8\ell_s^B,\chi=0.22$}
   \psfrag{d}[l]{$k_0\ell_s^B=18,L=11\ell_s^B,\chi=0.44$}
   \psfrag{e}[l]{$k_0\ell_s^B=36,L=2\ell_s^B,\chi=0.22$}
   \psfrag{f}[l]{$k_0\ell_s^B=36,L=2\ell_s^B,\chi=0.44$}
   \psfrag{g}[l]{$k_0\ell_s^B=36,L=4\ell_s^B,\chi=0.22$}
   \psfrag{h}[l]{$k_0\ell_s^B=36,L=5\ell_s^B,\chi=0.44$}
   \includegraphics[width=1\linewidth]{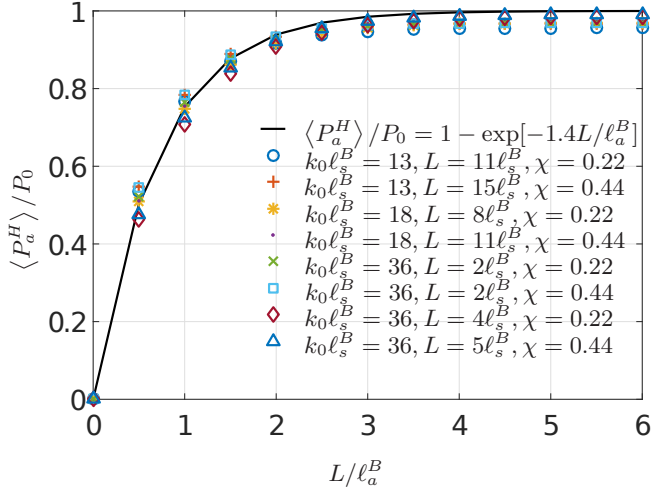}
   \caption{Normalized average absorbed power for stealth hyperuniform structures operating in the transparent
   conditions showing weak variability of the absorption mean free path $\ell_a^H$ with the correlations.}
   \label{fig:l_a_correlations}
\end{figure}

\Fig{l_a_correlations} shows the normalized average absorbed power of stealth hyperuniform clouds with different values
of the parameters of interest such as the ISA optical thickness $L/\ell_s^B$, the order degree $\chi$, or
disorder strength $k_0\ell_s^B$.

As the extinction length is only given by the absorption length $\ell_a^H$, the weak variability observed in the
curves shows that $\ell_a^H$ is weakly affected by the presence of structural correlations. This is shown here for
stealth hyperuniform media but it was already stressed for hard-spheres correlations~\cite{LESEUR-2016,WANG-2018}.

Nevertheless, a dummy fit by an exponential shows that the absorption mean free path deviates approximately by a factor
$\ell_a^H/\ell_a^B\simeq 1/1.4\simeq 0.7$ compared to the ISA absorption mean-free path and the normalized absorbed
power is not exactly unity for very large absorption optical thickness. This comes from the fact that small
corrections have to be taken into account due to the correlations in the scatterers positions, but also from the rather high density of
scatterers (chosen due to computational constraints) and index mismatch at the interface [as shown by \eq{P_a_hyp}].

For very dilute systems, the index mismatch can be neglected and assuming $\ell_a^H\sim\ell_a^B$, we obtain a
simplified fully analytical expression for the average power absorbed by the hyperuniform structure given by
\begin{equation}\label{eq:P_a_hyp_simplified}
   \frac{\bra P_a^H\ket}{P_0}=\left[1-\exp\left(-\frac{L}{\ell_a^B}\right)\right]
\end{equation}
which is the expression used to evaluate the average absorption power in the stealth hyperuniform case of
\fig{gain_estimate}. Note that the assumption $\ell_a^H\sim\ell_a^B$ is a convenient approximation but is not strictly
correct as specified above.

\section{Average absorbed power for the uncorrelated structure}\label{app7}

In this appendix, we derive an analytic expression of the average power absorbed by an uncorrelated cloud using the
diffusion approximation. Compared to Monte Carlo simulations, the derivation leads to a simple formula of the estimated
gain for very large optical thicknesses. The diffusion approximation is valid (1) whenever the
size of the cloud $L$ is much larger than the scattering mean-free path $\ell_s^B$, (2) under the weak absorption
condition $\ell_s^B\ll\ell_a^B$ and (3) in dilute systems such that $k_0\ell_s^B\gg 1$. The system is considered
diluted enough to neglect index mismatch at the interfaces and to set $\ell_s^U=\ell_s^B$ and
$\ell_a^U=\ell_a^B$.  Finally, as we consider uncorrelated point dipoles scatterers, the transport mean-free path is
$\ell^*=\ell_s^B$. The diffusion equation for the 2D slab geometry and for a plane-wave illumination at normal
incidence is given by~\cite{ISHIMARU-1997}
\begin{equation}
   \frac{c\ell_s^B}{2}\frac{\ud^2 \widetilde{u}}{\ud z^2}-\frac{c\widetilde{u}}{\ell_a^B}=-\frac{\overline{P}}{\ell_s^B}
\end{equation}
where $\widetilde{u}$ is the energy density of the diffuse part of the beam and
\begin{equation}
   \overline{P}(z)=P_0\exp\left(-\frac{z}{\ell_e^B}\right)
\end{equation}
is the ballistic source term. To take into account properly the boundary conditions, we use
\begin{align}
   \widetilde{u}(z=0)-z_0\frac{\ud \widetilde{u}}{\ud z}(z=0) & =0,
\\
   \widetilde{u}(z=L)+z_0\frac{\ud \widetilde{u}}{\ud z}(z=L) & =0,
\end{align}
where $z_0/\ell_s^B=\pi/4$ is the 2D extrapolation length~\cite{MONTAMBAUX-2007}. The solution of this set of
equations reads
\begin{equation}
   \widetilde{u}=\frac{P_0\gamma}{c}\left[\alpha^+\exp\left(\frac{z}{\ell}\right)+\alpha^-\exp\left(-\frac{z}{\ell}\right)
      -\exp\left(-\frac{z}{\ell_e^B}\right)\right]
\end{equation}
where
\begin{multline}
   \alpha^{\pm}=\left[\mp\left(1+\frac{z_0}{\ell_e^B}\right)\left(1\mp \frac{z_0}{\ell}\right)\exp\left(\mp\frac{L}{\ell}\right)
\right.\\\left.
                     \pm\left(1-\frac{z_0}{\ell_e^B}\right)\left(1\pm \frac{z_0}{\ell}\right)\exp\left(-\frac{L}{\ell_e^B}\right)\right]
\\\times
      \left[2\left(1+\frac{z_0^2}{\ell^2}\right)\sinh\left(\frac{L}{\ell}\right)+2\frac{z_0}{\ell}\cosh\left(\frac{L}{\ell}\right)\right]^{-1},
\end{multline}
\begin{equation}
   \gamma=\frac{2}{1+{\ell_s^B}^2/{\ell_a^B}^2}
\end{equation}
and $\ell=\sqrt{\ell_a^B\ell_s^B/2}$. The power absorbed by the slab is given by
\begin{equation}
   \bra P_a^U\ket=\frac{c}{\ell_a^B}\int_0^L\bra u(z)\ket\ud z
\end{equation}
where $\bra u(z)\ket=\overline{u}(z)+\widetilde{u}(z)$ is the average energy density. We have
$\overline{u}(z)=\overline{P}(z)/c$, thus we obtain
\begin{multline}\label{eq:P_a_uncorr}
   \frac{\bra P_a^U\ket}{P_0}=\frac{\ell_e^B}{\ell_a^B}\left[1-\gamma\right]\left[1-\exp\left(-\frac{L}{\ell_e^B}\right)\right]
\\
   +\frac{\ell}{\ell_a^B}\left\{\alpha^+\left[\exp\left(\frac{L}{\ell}\right)-1\right]
                                 +\alpha^-\left[1-\exp\left(-\frac{L}{\ell}\right)\right]\right\}
\end{multline}
which is the expression used to evaluate the average absorption power in the uncorrelated case of \fig{gain_estimate}.

\section{Estimation of the asymptotic maximum gain}\label{app8}

\begin{figure}[!htb]
   \centering
   \includegraphics[width=0.8\linewidth]{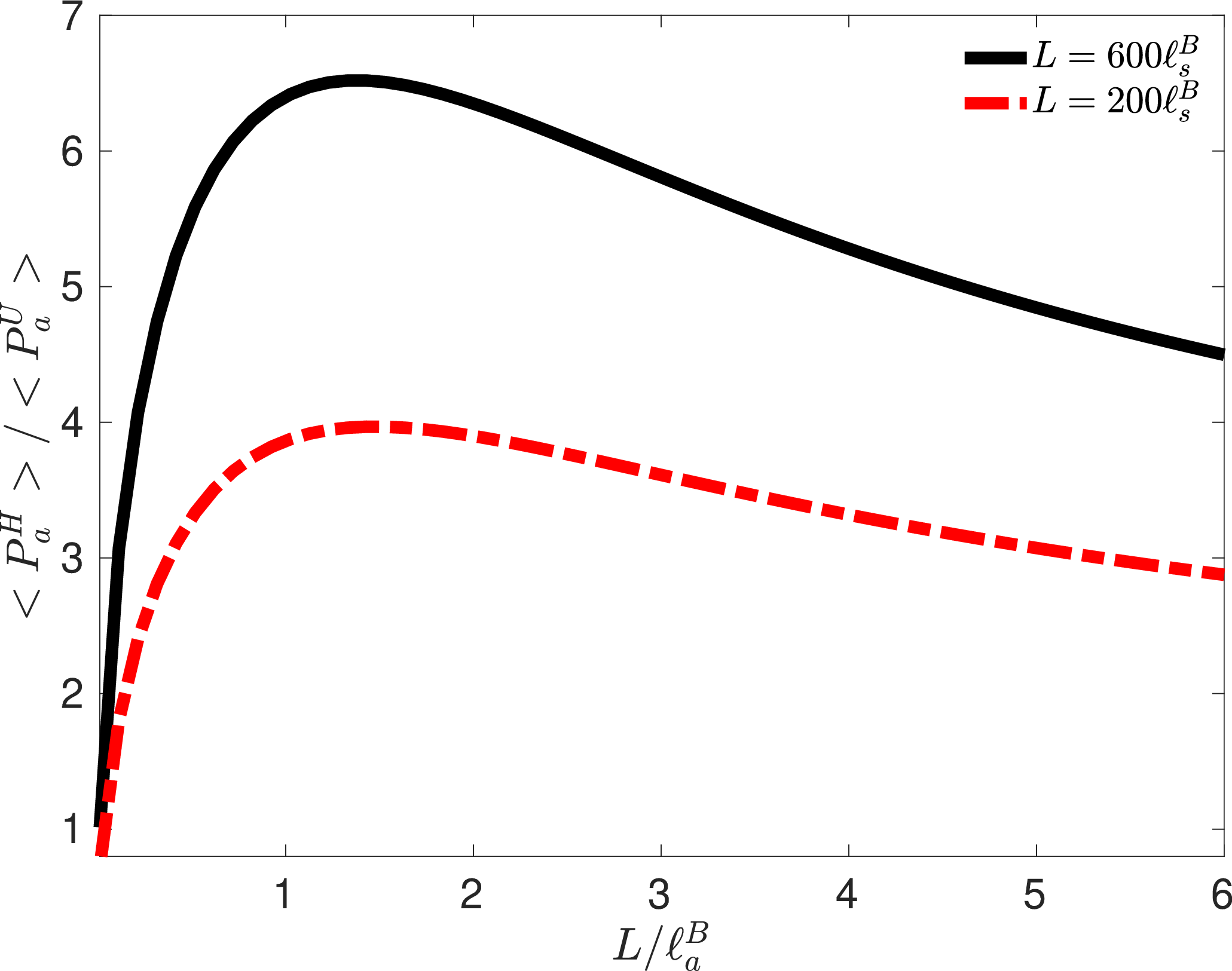}
   \caption{Estimated gain for large optical thicknesses.}
   \label{fig:gain_estimate}
\end{figure}

To estimate the asymptotic maximum gain that we can get, we use \eq{P_a_hyp_simplified} for the stealth hyperuniform
structure. Regarding the uncorrelated system, we need to simplify \eq{P_a_uncorr} under the conditions $L/\ell_e^B\gg
1$, $L/\ell\gg 1$. By keeping only the second order terms of the expansion in $\ell_s^B/\ell_a^B$ we obtain
\begin{multline}
   \frac{\bra P_a^U\ket}{P_0}=\left(1+\frac{z_0}{\ell_s^B}\right)\sqrt{\frac{2\ell_s^B}{\ell_a^B}}
      -\left[1+2\frac{z_0}{\ell_s^B}\left(1+\frac{z_0}{\ell_s^B}\right)\right]\frac{\ell_s^B}{\ell_a^B}
\\
      +\mathcal{O}\left(\frac{\ell_s^B}{\ell_a^B}\right).
\end{multline}
The second term may be used to estimate the relative error commited by keeping only the first term. For instance, for
$\ell_a^B=40\ell_s^B$, the relative error is approximately $\SI{25}{\%}$ and for $\ell_a^B=240\ell_s^B$, the relative
error is about $\SI{10}{\%}$. We then keep only the first term and we define the gain through the relation
\begin{equation}
   G\left(\frac{L}{\ell_s^B},\frac{L}{\ell_a^B}\right)
      =\frac{\bra P_a^H\ket}{\bra P_a^U\ket}.
\end{equation}
The maximum gain is obtained when
\begin{equation}
   \frac{\partial G}{\partial (L/\ell_a^B)}=0
\end{equation}
which leads to
\begin{align}\label{eq:max}
   \left[\frac{L}{\ell_a^B}\right]_{\text{optimum}} & =1.256,
\\
   \max\left[\frac{\bra P_a^H\ket}{\bra P_a^U\ket}\right]
      & \sim 0.243\sqrt{\frac{L}{\ell_s^B}}.
\end{align}
From this expression, we predict a gain of the order of $4$ for $L=200\ell_s^B$ and $7$ for $L=600\ell_s^B$ (see
\fig{gain_estimate}). It is important to note that \eq{max} gives only an order of magnitude of the
maximum gain and refinements are needed to be more accurate by taking into account the real effective refractive index
for both the hyperuniform and the uncorrelated structures and the multiple scattering process beyond the diffusion
approximation.

%

\end{document}